# SHIELD: Sustainable Hybrid Evolutionary Learning Framework for Carbon, Wastewater, and Energy-Aware Data Center Management

Sirui Qi
*Colorado State University*
Fort Collins, USA
alex.qi@colostate.edu

Dejan Milojicic, Cullen Bash
*Hewlett Packard Labs*
Milpitas, USA
{dejan.milojicic, cullen.bash}@hpe.com

Sudeep Pasricha
*Colorado State University*
Fort Collins, USA
sudeep@colostate.edu

*Abstract—* **Today's cloud data centers are often distributed geographically to provide robust data services. But these geo-distributed data centers (GDDCs) have a significant associated environmental impact due to their increasing carbon emissions and water usage, which needs to be curtailed. Moreover, the energy costs of operating these data centers continue to rise. This paper proposes a novel framework to co-optimize carbon emissions, water footprint, and energy costs of GDDCs, using a hybrid workload management framework called SHIELD that integrates machine learning guided local search with a decomposition-based evolutionary algorithm. Our framework considers geographical factors and time-based differences in power generation/use, costs, and environmental impacts to intelligently manage workload distribution across GDDCs and data center operation. Experimental results show that SHIELD can realize 34.4× speedup and 2.1× improvement in Pareto Hypervolume while reducing the carbon footprint by up to 3.7×, water footprint by up to 1.8×, energy costs by up to 1.3×, and a cumulative improvement across all objectives (carbon, water, cost) of up to 4.8× compared to the state-of-the-art.**

## I. Introduction

In recent years, the emerging use of general-purpose chat-bots, recommendation engines, and Internet-of-Things (IoT) devices [1] has increased the reliance on cloud data centers. Cloud service providers have been gradually distributing their data centers geographically across multiple locations. Such geo-distributed data centers (GDDCs) have many advantages. Establishing data centers closer to large customer bases offers better performance and lower network costs for them. Multiple data centers also provide better resilience to catastrophic failures (e.g., environmental hazards) that may impact a data center at a specific location.

However, thriving GDDCs are exacerbating the energy consumption and environmental impacts of cloud computing all over the world. Today, data centers account for 1% of worldwide electricity usage [2] and 0.6% of global greenhouse gas emissions [3]. There are over 2600 data centers in the United States and nearly 8000 data centers across the world, and this number is projected to increase in the coming decade [4]. These data centers consume large quantities of water, e.g., Google's 14 data centers consumed 4.3 billion gallons of water in 2021 [5], which puts immense pressure on local water supplies. Due to global climate change and the tightened energy policies in many nations [6], [7], researchers have recognized the need for realizing sustainable data centers.

Reducing the energy costs and environmental (water, carbon) overheads of GDDCs has thus taken on great urgency. From the perspective of minimizing energy costs, workloads should be assigned to data centers where there is cheap energy. Meanwhile, from the perspective of improving sustainability, workloads should be assigned to data centers that can provide cleaner (e.g., solar, wind) energy sources. These two perspectives are usually in conflict, and it is the cloud service providers responsibility to manage workloads judiciously, so that both energy cost and sustainability goals can be met at the same time.

GDDCs provide compelling opportunities to better manage energy costs and environmental impacts. For example, exploiting time-of-use (TOU) electricity pricing [8] can allow workloads to be executed at GDDC locations with lower TOU pricing (e.g., during off-peak periods), to reduce energy costs. Compared with peak prices, the off-peak price can be 10× lower. Another opportunity is to utilize green energy techniques such as free air cooling [9] which may be available at some locations. By utilizing direct air exchangers instead of mechanical refrigeration components (e.g., air conditioner), free air cooling can reduce both environmental impacts and energy costs in data centers. However, effective free air cooling has maximal outdoor temperature and dew point constraints according to ASHRAE (American Society of Heating, Refrigerating and Air-Conditioning Engineers) [10]. Temperature and dew point values also vary with time and location. Thus, to optimize GDDC operation, an effective GDDC management policy must consider time- and geography-based differences across data centers and exploit them carefully.

Prior efforts to address the GDDC management problem have focused on either minimizing energy costs (e.g., [11]) or an isolated sustainable goal such as carbon minimization or water-use minimization (e.g., [12], [13]). However, from a cloud service provider's perspective, there is a need to simultaneously optimize for all of these goals. Such multi-objective (energy, carbon, water) co-optimization is a complex problem for many reasons. For instance, energy consumption reduction in GDDCs by migrating a workload to an energy cost-efficient data center may lead to increased carbon and water usage as the chosen data center might rely on a more carbon-intensive energy source. Meanwhile, excessive pursuit of an isolated sustainable goal (e.g., carbon emission reduction) may negatively impact another sustainable goal (e.g., causing an increase in water use) or an unacceptable rise in energy costs.

To address these important challenges, in this work, we propose a novel and efficient multi-objective optimization framework to co-optimize carbon emissions, water footprint, and energy costs of GDDCs. Our proposed S̲ustainable Hy̲br̲id E̲volutionary L̲earning Framework for Geo-Distributed D̲ata Center Management (SHIELD) performs intelligent design space exploration to generate efficient Pareto-optimal solutions that minimize the energy costs and environmental impact of GDDCs. The novel contributions of our work can be summarized as follows:

- We comprehensively model the carbon emissions, water profile, and energy use of GDDCs;
- We formulate a three-objective optimization problem for sustainable GDDC operation which involves minimizing carbon emissions, water footprint, and energy costs;
- We propose a new framework called SHIELD that combines machine learning and evolutionary algorithmic techniques to co-optimize the three objectives for GDDCs;
- We compare SHIELD with the state-of-the-art data center management frameworks and show that SHIELD outperforms them in terms of Pareto Hypervolume, convergence speed, scalability, and solution quality.

The rest of the paper is organized as follows. In Section II, we review relevant prior works. We characterize our models in Section III. Sections IV and V describe our problem formulation and the proposed SHIELD framework. The comparison methodology and experiment results are presented in Section VI. Lastly, we present concluding remarks in Section VII.

## II. RELATED WORK

Cloud resource management has been studied for many years [14]. Single-objective challenges in cloud management such as quality of service [15], cost [16], and resource utilization rate [17] have been addressed by different methods. Several multi-objective optimization techniques have also been applied to cloud management in recent years, including simulated annealing (SA), genetic algorithm (GA), and non-dominated sorting [11], [18], [19].

Liu et al. proposed a holistic optimization framework for mobile cloud workload computing [18]. Their framework focused on triple-objective optimization (TOO) of energy consumption, system reliability, and quality of service. Their framework used SA, which is a probabilistic approach to approximate the global optimum by mimicking the slow cooling of metal. SA was shown to be effective in providing trade-offs across the three objectives. However, SA does not have obvious advantages in performance and speed for larger design space exploration compared to other approaches, such as GA.

GAs are adaptive heuristic search algorithms inspired by natural selection and have been shown to be efficient for many optimization problems including workload scheduling. Hogade et al. proposed the genetic algorithm load distribution (GALD) approach to optimize energy costs in GDDCs [11]. This framework explored the potential of GDDCs, such as TOU electricity price, peak shaving, net metering, and local renewable energy availability. The consideration of these factors in the GA helped optimize energy costs. However, the authors only performed experiments on a single objective (energy cost). GALD can be easily extended to a multi-objective problem [20], but the main concern for GAs is slow convergence rates [21].

Bi et al. proposed a decomposition-based multi-objective evolutionary algorithm with Gaussian mutation and crowding distance (DMGC), which co-optimized energy cost and revenue of workload scheduling in data centers [19]. Compared with GA, they were able to preserve more diverse designs in their solution set. The diversity in the population further benefited subsequent mutations and led to better final solutions. At the same time, Gaussian mutation helped DMGC to jump out of local optima and converge to better solutions more quickly. In Gaussian mutation, the random offspring generation follows the Gaussian distribution pattern, taking parent solutions as the center of the distribution. However, our analysis indicates that the link between crowding distance and solution quality is weak. Crowding distance can filter out some high-quality designs and may also select designs in unneeded directions. The Gaussian mutation appears to be more suitable to problems where there are not too many constraints on design space, which is not the case with GDDC management.

Our proposed framework (SHIELD) combines machine learning and evolutionary algorithms to overcome many of the drawbacks of state-of-the-art frameworks for data center management. Further, we tackle a more complex multi-objective optimization problem and use a more comprehensive system model than prior works in this area.

## III. SYSTEM MODEL

Our framework involves the mapping of each workload from its origin to different geographic locations and then to different nodes inside a data center. We further characterize energy costs and the environmental impact triggered by each mapping. Meanwhile, our framework can co-optimize the energy cost and environmental impacts by adjusting mappings and corresponding energy payment plans. Fig. 1 illustrates a data center's air/water/electricity flow that is modeled in our framework. Each data center is comprehensively modeled in terms of its power use, carbon emissions, water use, energy costs, and workload, as discussed in the rest of this section.

### A. Power Model

#### 1) Data Center Layout

Each data center consists of $N$ computing nodes, arranged in rows of racks [22]. The racks are arranged in a standard hot-aisle/cold-aisle configuration [23]. We assume several computing node types in data centers, whose number also varies across locations. These nodes also have different energy profiles for different computing workloads. Such heterogeneity in the GDDC node configurations is becoming widespread due to diverse workloads and service level agreements. CRAC units are used to cool the data center room. Besides mechanical cooling systems, we consider free air cooling availability in a subset of data centers. Free air cooling involves pumping cold outdoor air directly into the data center through direct air exchangers. We assume that these direct air exchangers are equipped with high-performance filters [24] to resist air pollution. When the outdoor temperature and dew point both allow, the data center will switch to free air cooling from mechanical cooling [9].

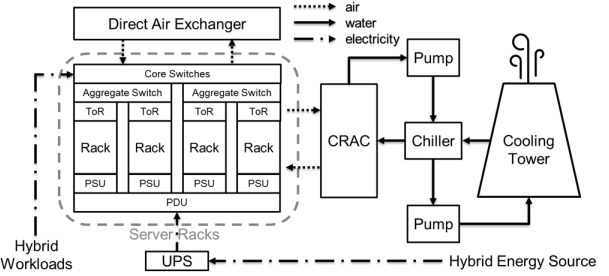

Fig. 1. Air/Water/Electricity flow in a data center. Top-of-rack (ToR) switch is part of the switch system to provide server-to-server connections. Power supply unit (PSU), power distribution unit (PDU), and uninterrupted power supply (UPS) constitute the power switching system. Computer room air conditioning (CRAC) units are a part of the cooling system.

#### 2) Power Components

We divide the power consumption of a data center $i$ into information technology (IT) load $P_{IT,i}$, cooling $P_{Cooling,i}$, and Internal Power Conditioning System (IPCS) $P_{IPCS,i}$ [25].

IT load tracks the power consumption of servers in the data center and is related to the executed workload. Consider $t$ different workload types computed by $a$ active nodes. We assume a fixed p-state for each node. The power consumption of the IT load can be calculated by Eq. (1) below, where $AP$ represents active power of workload type $i$ assigned at node $j$, and $IP$ is idle power of node $k$.

$$P_{IT} = \sum_i^a \sum_j^t AP_{ij} + \sum_k^{N-a} IP_k \qquad (1)$$

There are three components of cooling power: CRAC units, chillers, and supporting equipment such as cooling tower, pumps, etc. [26]. From [22], the coefficient of performance ($CoP$) is defined as the ratio of removed heat to the amount of necessary power to remove the heat. The CRAC power consumption can be estimated as:

$$P_{CRAC} = P_{IT} / CoP \qquad (2)$$

From [26], the power consumption of chillers and the supporting equipment is close to the power consumption of CRAC units. Hence, we can estimate the whole cooling power as:

$$P_{Cooling} = 3 \times P_{CRAC} = 3 \times P_{IT} / CoP \qquad (3)$$

The IPCS comprises power management components such as PDU, PSU, and UPS. The power consumption of IPCS correlates to the IT load and can be estimated as [25]:

$$P_{IPCS} = 0.13 \times P_{IT} \qquad (4)$$

### B. Water Model

We consider both site-based (from cooling) and source-based (from electricity generation) water consumption for data centers. By doing so, we can evaluate how geography-based differences impact water consumption. The overall water footprint of $D$ GDDCs can be calculated by Eq. (5) below, in which $V_{E,i}$, $V_{B,i}$, and $V_{S,i}$ are volumes of evaporative, blowdown-to-wastewater-facility, and source water consumption of data center $i$ respectively:

$$V_{ALL} = \sum_i^D (V_{E,i} + V_{B,i} + V_{S,i}) \qquad (5)$$

Direct water consumption is common in mechanical cooling data centers, where the water is used as a coolant. Incoming water to data centers is usually potable water from water plants, while outgoing water from data centers is considered industrial wastewater, which needs to be processed at a wastewater facility [27]. We estimate the volume of direct water consumption through all means of water outflows, which are evaporative water through the cooling tower and blowdown water to the wastewater facility.

The evaporative water consumption $V_E$ through a cooling tower can be calculated by Eq. (6) below. $E_{IT}$ represents the heat generated by IT infrastructure and $H_{water}$ is the latent heat of the water.

$$V_E = E_{IT}/H_{water} \quad (6)$$

The second part of site-based water outflow is the volume of blowdown water $V_B$ to the wastewater treatment facilities. We assume all data centers cycle potable water until the concentration of dissolved solids is roughly $C$ times the supplied water [28]. Hence, we estimate the volume of blowdown water by Eq. (7):

$$V_B = V_E/(C-1) \quad (7)$$

Source-based water consumption $V_S$ is primarily from electricity generation which utilizes brown energy sources such as coal. Modern power grids usually utilize different energy sources and their brown energy ratios vary across locations. For example, the energy water intensity factor ($EWIF$) in Maryland is 0 while in Illinois it is 3.97 L/kWh [29], i.e., 3.97 liters of water are used when generating 1 unit of electricity in Illinois. We can calculate source-based water consumption based on energy $E$ consumed at data center as:

$$V_S = E \times EWIF \quad (8)$$

### C. Carbon Model

Carbon dioxide is one of the biggest sources of greenhouse gas emissions [30]. Prior efforts on data center carbon emission reduction solely consider the minimization of electricity-based carbon emissions and ignore water-use-based carbon emissions. In this work, one of our novel contributions is to find correlations over carbon, water, and energy use in data centers, and co-optimize these. We thus analyze the carbon footprint from not just electricity generation but also potable water usage and wastewater treatment. Our analysis reveals that data centers with mechanical cooling may have a larger carbon footprint than expected. The overall carbon emission of $D$ GDDCs can be calculated by Eq. (9) below, where $M_{electricity,i}$ and $M_{water,i}$ are the mass of electricity-based and water-based carbon emitted at the data center $i$ respectively:

$$M_{ALL} = \sum_i^D (M_{electricity,i} + M_{water,i}) \quad (9)$$

As geographical differences introduce energy source differences when calculating the electricity-based carbon emission, Carbon Factor $CF$ measures the mass of emitted carbon during the process of electricity generation. We use the geographical $CF$ from [31], and formulate the estimation function in Eq. (10), in which $M_{electricity}$ represents the mass of electricity-based carbon emissions and $E_B$ is the amount of brown energy used in a data center:

$$M_{electricity} = E_B/CF \quad (10)$$

We also characterize water-based carbon emissions due to the production of potable water and the treatment of wastewater. The water plant and wastewater treatment facility are assumed to use electricity from the local power grid. By combining geographical $EWIF$ [32] and $CF$, we obtain the water-based carbon emission from Eq. (11) below where $I_P$ and $I_W$ are the energy intensities for potable water production and wastewater treatment (representing the energy consumption per unit of water treatment):

$$M_{water} = [(V_B + V_E) \times I_P + V_S \times I_W]/CF \quad (11)$$

### D. Energy Cost Model

We consider three price models that are relevant to estimating the energy costs associated with distributing workloads across GDDCs: (a) TOU price, (b) clean premium, and (c) annual clean contract.

As discussed earlier, TOU price is a key factor that can help determine when to shift workloads to off-peak periods at different time zones, to reduce the energy cost of computing. However, this can lead to high environmental impacts because cheap power may not always be green power. For example, off-peak periods occur usually at midnight when there is no solar energy available.

To further enable trade-offs between energy cost and environmental impacts, we consider two additional factors: clean premium and annual clean contract.

Clean premium is an extra fee that cloud service providers pay for clean energy use from the power grid. Once the extra fee is paid over the original TOU electricity price, the power provider can provide electricity from renewable energy sources. The premium price model is already available in many local power markers such as San Francisco and Denver. Besides the reduction in environmental impacts, the clean premium also stands out for its configurability. For example, consider a data center that consumes two units of energy at a local power market. Cloud managers can apply a premium on one unit of energy (zero carbon emission), with the other unit of energy derived from mixed energy sources. By doing so, the cloud service provider can balance energy costs and environmental impacts.

Another cheaper clean energy price model is the annual clean contract, which only exists in the Texas area in the United States. Several power providers such as Gexa Energy in Texas provide 24-hour all-year-around electricity from renewable sources. Compared with a clean premium, an annual clean contract can be cheaper because if the annual energy use of a data center can be estimated in advance, the cloud manager can then sign an unchangeable annual contract and get cheaper clean energy than with a clean premium.

Our framework has the ability to judiciously balance TOU prices, clean premiums, and annual clean contracts when distributing workloads across GDDCs. The geography- and time-based differences in TOU prices and annual clean contracts will influence our framework's distribution decision on every workload. Meanwhile, the framework needs to self-decide the amount of clean premiums at every data center location to co-optimize its objectives.

### E. Workload

We consider a rate-based workload management scheme, where the workload arrival rate can be estimated over a decision interval called an epoch [33]. In our work, epoch length is one hour, and thus a 24-epoch period represents a full day. Within the short duration of each epoch, workload arrival rates can be reasonably approximated as constant [34]. As shown in Eq. (12), our GDDC management framework needs to map the global arrival rate $GAR_j$ of workload $j$ into local arrival rates $AR_{i,d}$ across the $D$ GDDCs:

$$GAR_j = \sum_i^D AR_{i,j} \quad (12)$$

The following subsections present our problem formulation and describe our SHIELD framework for GDDC management.

## IV. PROBLEM FORMULATION

We consider a cloud service provider managing GDDCs across multiple locations across the USA. A GDDC management framework must distribute the workload coming in from various locations to data centers in the cloud service providers' GDDC. In each epoch (hour), the framework is responsible for providing distribution plans that include two parts: *(i)* workloads assigned to each location, and *(ii)* the amount of clean premium at each location. The goal of the framework is to co-optimize three objectives: energy cost, carbon emissions, and water footprint. In our initial assumptions, the GDDCs are under-subscribed in the sense that they are expected to have enough computation resources to prevent any workload from being dropped or terminated before completion. The workloads originate off-site from the data centers, and we consider only workloads with negligible transfer time and costs. Once workloads are assigned to a data center, the same local data center scheduling policy is used to schedule local workloads no matter the location. The local scheduling policy is primarily based on workload type and node type. It builds on the list scheduling approach, where an ordered list of available heterogeneous nodes based on execution times is maintained for each workload type, to guide the mapping [35].

## V. SHIELD FRAMEWORK

Our proposed SHIELD framework integrates a novel hybrid search approach that utilizes Machine Learning (ML)-guided local search with priorities and a Decomposition-based Evolutionary Algorithm (EA) with Knowledge Propagation. As shown in Fig. 2, a randomly generated population is input to an ML module for local search starting point selection. Due to a lack of training data in early iterations, the ML module randomly picks starting points in the beginning. After some time, these points are locally searched by the local search model based on improvement in their weighted sums. After local search, the population is updated with local search results and their trajectories are stored for ML module training. Our EA model further explores the design space and helps the local search model jump out of its local optima. After our EA updates the population, a new iteration starts with the ML module selecting the starting points. Algorithm 1 summarizes the pseudo-code for SHIELD, which includes ML-guided local search (lines 2-7) and decomposition-based EA with knowledge propagation (lines 8-9). The input includes maximal generations to consider for optimization $gen$, population size $N$, number of objectives $M$, number of early iterations for random local search $iter_{early}$, and ML module update frequency $f_{update}$. The output consists of $N$ Pareto optimal designs for $M$ objectives. The objective values of each design $p$ are calculated based on models introduced in Section IV. Details of decomposition-based EA with knowledge propagation and ML-guided local search are discussed in the following subsections.

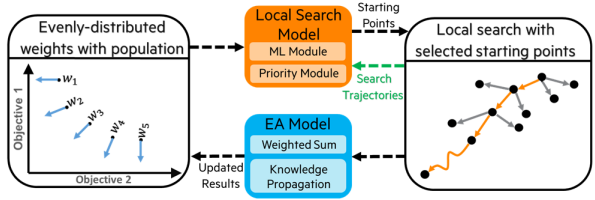

Fig. 2 SHIELD framework overview

---

**ALGORITHM 1**: SHIELD framework

**Input:** $gen, N, M, iter_{early}, f_{update}$
**Output:** Population $P$ (Final N designs)
**Initialization:**
**Evenly Distributed Weight Vector Set** $W = \{w_1, ..., w_N\}$
**Randomly Generated Population** $P = \{p_1, ..., p_N\}$
**Training Set for ML Module** $S_{train} \leftarrow \emptyset$
**Ideal Point in *M*-objective Design Space** $z = [o_1, ..., o_M]$

1: **for** $i = 0$ to $gen$ **do**
2:    **if** $i < iter_{early}$: $P_{start} \leftarrow Random(n_{local}, P)$
3:    **else if** $(i - iter_{early}) \% f_{update} == 0$:
       $Eval \leftarrow MLtrain(S_{train}); S_{train} \leftarrow \emptyset$
4:    **else** $P_{start} \leftarrow MLguide(eval, W, P); P_{start} = P_{start} \cup P_{priority}$
5:    $P_{new} \leftarrow \emptyset$
6:    **for** $p, w$ in $P_{start}$ **do**
       $p_{new}, S_{trsj} = LocalSearch(w, p); P_{new} = P_{new} \cup \{p_{new}\}$
       $S_{train} = S_{train} \cup S_{traj}$
   **end for**
7:    $P = (P - P_{start}) \cup P_{new}$
8:    $P_{rest} = P - P_{start}; P_{offspring} = EA(P_{start}, P_{rest})$
9:    **for** p in $P_{offspring}$ **do** $P \leftarrow Update(P, p, W)$ **end for**
10: **end for; return** $P$

---

*A. Decomposition-based EA with Knowledge Propagation*

Unlike decomposition-based EA such as [36] which performs crossover and mutation only in a neighboring local space, our EA model realizes knowledge propagation by purposefully performing crossover across locally-searched points $P_{start}$ and non-locally-searched points $P_{rest}$ (line 8). Subsequently, our model mutates the crossover offspring to further explore the design space. Thus, our EA model guides crossover and mutation between locally-searched points and non-locally-searched ones, with knowledge gained during local search being propagated to non-locally-searched points, which avoids performing a local search on the whole population while still improving the overall population quality by expanding the exploration space. The generated offspring are used to update the population (line 9) via the function $Update(P, p, W)$, where each design point $p$ in offspring $P_{offspring}$ is randomly compared with design points in $P$ with the weighted sum function:

$$minimize\ g(x|w, z) = \Sigma_{i=1}^{M}\{w_i|Obj_i(x) - z_i|\} \quad (13)$$

This weighted sum approach is different from the decomposition method used in existing decomposition-based EAs, which involve the Tchebycheff approach [37] and use a set of $N$ uniformly spread weight vectors $W = \{w_1, ..., w_N\}$ in the following manner:

$$minimize\ g(x|w, z) = \max_{1 \leq i \leq M}\{w_i|Obj_i(x) - z_i|\} \quad (14)$$

where $g$ is the scalar optimization problem, $M$ is the number of objectives, $Obj_i(x)$ is the $i^{th}$ objective value of input $x$, and $Z = \{z_1, ..., z_m\}$ is the ideal point defined as the minimum value of all the objectives. Given a weight vector $w_i$, a lower Tchebycheff value $g(x|w, z)$ means a better design is found for the $i^{th}$ subproblem. Instead of this approach, the weighted sum approach from Eq. (13) is deployed in our update function (line 9), to help our EA explore the design space. Compared with [37], our weighted sum approach can provide more diverse and fine-grained optimization directions (see results in Section VI) during design space traversal.

*B. ML-guided Local Search with Priority*

To boost decomposition-based EA with knowledge propagation both in speed and quality, a local search model is introduced in our framework. Specifically, an ML module and priority module are integrated into our local search model, to guide the local search. We observed that a local search model not only speeds up the convergence, but also can potentially provide much better individuals for EA to select as parents and then generate better offspring. These outstanding individuals, when generated by a local search model, will be much better than other individuals in the same generation. We have found that even a few outstanding individuals in a huge population can lead to a much better next-generation population when utilized by the same EA. Meanwhile, inspired by STAGE [38] which selects local search starting points using an evaluation function, an ML module is integrated into this framework to predict local search results (weighted sum) by studying previous local search trajectories (visited designs and corresponding weight vectors). Furthermore, we use a greedy descent approach in local search based on the same weighted sum as EA does. The same update methodology can build a bridge between local search and EA, to ensure that learning is consistent across both.

The local search model starts with a random local search (line 2) to create a training dataset $S_{train}$ for subsequent ML module training $MLtrain(S_{train})$. The dataset consists of local search trajectories and is recreated at the frequency $f_{update}$ after being used for ML module training (line 3). By using $f_{update}$, not only can we reduce the time redundancy introduced by ML module training, but we also keep $S_{train}$ updated and compact. The ML module we use is a random forest model, which is an ensemble model that uses the average output from a collection of decision trees to help reduce overfitting. The evaluation function $eval$ of the module maps each design's parameters and weight to the result of the search (Eq. (13)).

After $iter_{early}$ iterations, the local search model performs $eval$ on all design points in $P$ and selects local search starting points $P_{start}$ based on predicted weighted sum improvement over the current weighted sums of $P$. This starting point selection process is represented by $MLguide(eval, W, P)$ in line 4. In this manner, we select local search starting points with the most potential for improvement. Meanwhile, a set of priority design points $P_{priority}$ is added to $P_{start}$ even though they have a smaller predicted weighted sum improvement than the original $P_{start}$. $P_{priority}$ is an enhancement to force local search in desired directions. These directions are usually single objective-efficient and have less weighted sum improvement in local searches. However, we find that searches on single objective-efficient directions can better explore the design space and improve the quality of non-locally-searched points $P_{rest}$ through the EA model.

At each generation, an empty set $P_{new}$ is used to store all endpoints $p_{new}$ from local search (line 5). Each point $p$ in $P_{start}$ is then input to the local search function $LocalSearch(w, p)$ and the function returns the search endpoint $p_{new}$ and search trajectory $S_{traj}$ which is recorded in the training set $S_{train}$ that is aggregated over generations (line 6). Lastly, all local search endpoints replace their corresponding starting points in $P$ (line 7) to create an enhanced population that improves the outcomes from our EA approach.

## VI. EXPERIMENTS

### A. Experiment Setup

We compare our proposed SHIELD with three state-of-the-art approaches: simulated annealing-based tri-objective optimization (TOO) [18], genetic algorithm-based load distribution (GALD) [11], and decomposition-based multi-objective evolutionary algorithm with Gaussian mutation and crowding distance (DMGC) [19]. We extend these frameworks to our multi-objective problem to co-optimize energy cost, carbon emissions, and water consumption, by distributing workloads geographically. The design space exploration algorithm in SHIELD uses the following parameter values: $N = 30$, $gen = \infty$, $f_{update} = 50$, and $iter_{early} = 500$. All data centers use three different types of Intel server nodes in our experiments: E3-1225v3, E5649, and E5-2697v2. These three nodes differ in their number of cores, frequency, power profile, and memory. These three types of Intel nodes constitute 4320 computing nodes in each data center and their mix differs across locations. Due to the popularity of data analytics workloads among cloud service providers, we use data-intensive workloads from the BigDataBench 5.0 [27], as shown in Table I. We list other relevant parameters for our experiments in Table II. We consider 16 different data center locations, with diverse characteristics, as shown in Fig. 3. We evaluate all frameworks by determining their energy costs, carbon emissions, and water usage, as well as the Pareto Hypervolume (PHV) [39] of solutions.

TABLE I. WORKLOADS USED IN THE WORKLOAD MODEL

| Benchmark | Workload Types | Dataset | Size |
|---|---|---|---|
| LDA | offline analytics | Wikipedia entries | 233MB |
| K-means | offline analytics | Facebook network | 650MB |
| Naive Bayes | offline analytics | Amazon reviews | 500MB |
| Image-to-Text | artificial intelligence | Microsoft COCO | 600MB |
| Image-to-Image | artificial intelligence | Cityscapes | 100MB |

TABLE II. PARAMETERS USED IN EXPERIMENTS

| Parameter | Range |
|---|---|
| $CoP$ | $3.75 \sim 5.72$ [22] |
| $EWIF$ | $0 \sim 3.97\ L/kWh$ [29] |
| Water Latent Heat | $0.66\ kWh/L$ (at 40°C) |
| $I_P, I_W$ | $550kWh/ML, 640kWh/ML$ [32] |
| $CF$ | $99.7 \sim 775g/kWh$ [31] |
| TOU | $1.8 \sim 48¢/kWh$ |
| Clean Premium | $0.39 \sim 144¢/kWh$ |
| Annual Clean Contract | $15¢/kWh$ |
| Concentration Cycle ($C$) | 5 |

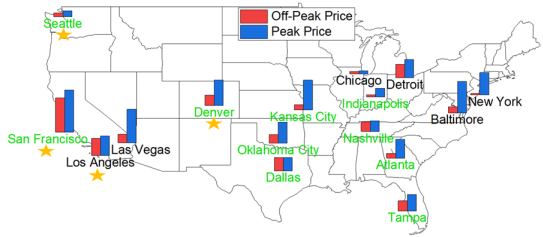

Fig. 3. Data center power price map with different price models: TOU price, clean premium, and annual clean contract. The red/blue bars indicate the local off-peak/peak power price, which is $1.8\sim48\ ¢/kWh$. Locations in green are ones with clean premium projects available. The starred locations are equipped with air-free cooling techniques which switch to air-free cooling in lower-than-75 ℉ temperatures and lower-than-63 ℉ dew points.

### B. Experiment Results

#### 1) PHV Improvements

To compare the quality of solutions generated by each framework, we first determined the PHV of generated solutions over time for a scenario with 16 data centers and all three objectives (minimizing energy cost, carbon footprint, and water footprint). PHV measures the size of the space enclosed by all solutions on the Pareto front and a user-defined reference point. A higher value of PHV is indicative of a more diverse and higher-quality solution set.

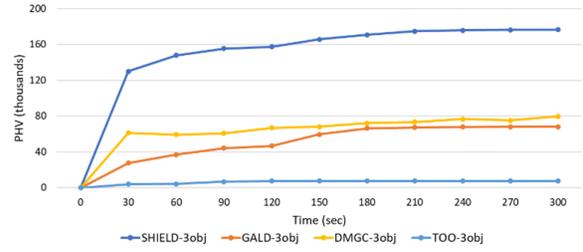

Fig. 4. Three Objective (energy cost, carbon footprint, water footprint) PHV over time of SHIELD, GALD, DMGC, and TOO.

From Fig. 4, we can observe that all frameworks converge within ~3 minutes. The PHV of SHIELD is 2.1× larger than that of the second-best framework (GALD). Further, to reach the second-best PHV result, it takes SHIELD 34.4× less time than GALD. Based on these results, we can see that SHIELD optimizes the PHV better and faster than other design space exploration frameworks.

Fig. 5 shows the corresponding 2D Pareto fronts from the PHV improvement experiment. It can be observed that SHIELD generates Pareto fronts that are more diverse and that contain solution configurations that dominate those generated from other frameworks.

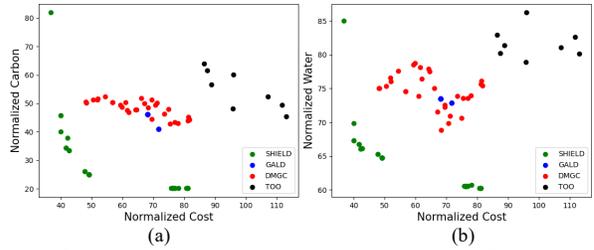

Fig. 5. Pareto front comparison of comparison frameworks for: (a) carbon emissions vs. energy cost; (b) water use vs energy cost

#### 2) Solution Quality Analysis

In our next experiment, we analyzed solution quality across our three metrics for a scenario with 16 data centers and a 1-minute runtime constraint, to ensure real-time decision making at the beginning of each epoch. Fig. 6 shows results aggregated over a 24-hour interval, with the y-axis showing improvements compared to the TOO framework. The three best solutions are selected from each framework. The most energy cost-efficient solution is depicted with the first three sets of bars that show the energy cost, carbon emissions, and water use of the most energy cost-efficient solution selected from each of the four compared frameworks. Similarly, the three middle sets of bars characterize the most carbon-efficient solution generated by the frameworks, while the last three sets of bars characterize the most water-efficient solution.

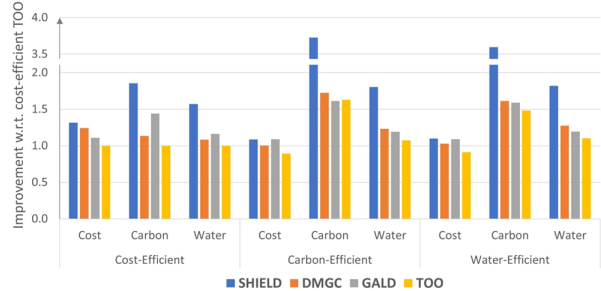

Fig. 6. 24-hour aggregated results for energy cost-, carbon-, and water-efficient solutions generated across the four frameworks.

From Fig. 6, it can be observed that SHIELD is always the best in cost/carbon/water reduction no matter which of the three metrics is prioritized. For the most energy cost-efficient solution (first three sets of bars), SHIELD has higher cost reduction as well as lower carbon emission and water consumption compared to all other frameworks. Similarly, for the most carbon-efficient solution (middle three sets of bars) and the most water-efficient solution (last three sets of bars), SHIELD generated higher quality solutions that outperform those from other frameworks. SHIELD reduces energy costs, carbon emissions, and water usage by up to 1.3×, 3.7×, and 1.8× respectively. We also determine a single best solution with the lowest cumulative energy cost, carbon emission, and water usage for each epoch, for each framework. Over a 24-hour interval, we found that SHIELD improves cumulative solution quality by 4.8×, 2.4×, and 3.2× when compared to TOO, GALD, and DMGC respectively.

*3) Scalability Analysis*

Lastly, we analyzed the scalability of the compared frameworks across different design spaces. We changed the size and complexity of the design space by modifying the number of data centers (Fig. 7(a)) and a number of objectives to co-optimize (Fig. 7(b)). The average PHV of the TOO framework is used as the baseline and we normalize the PHVs of other frameworks to it.

In Fig. 7 (a), we contrast PHV for three scenarios with 4, 8, and 16 data centers of our optimization problem. SHIELD outperforms other frameworks and interestingly, its benefits increase as the problem design space and complexity increases (from 4 to 16 data centers). For the 16 data center scenario, SHIELD's PHV is 2.1× higher than the second-best framework. This shows that SHIELD has good scalability with increasing data center size. From Fig. 7 (b), we can observe a similar phenomenon. As the number of objectives increases from one (1obj; energy cost) to two (2obj; energy cost and carbon emissions), and three (3obj; energy cost, carbon emissions, water use), SHIELD shows the best PHV improvement.

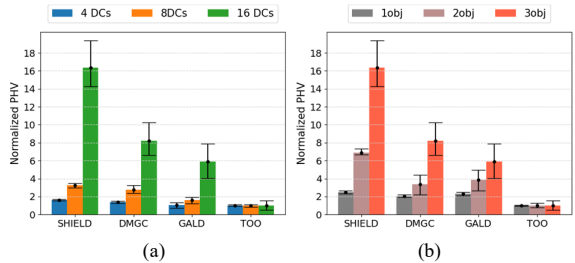

Fig. 7. Sensitivity analysis on: (a) number of DCs and (b) number of objectives. All PHV values are normalized w.r.t. the TOO framework. Error bars show the maximal/minimal PHV from 10 repeated experiments.

## VII. CONCLUSION

In this work, we studied the problem of workload distribution across geo-distributed data centers (GDDCs) to minimize energy cost, carbon footprint, and water use, simultaneously. We developed comprehensive models of energy consumption, energy price, water consumption, carbon emission, and workload execution. We then developed a novel framework called SHIELD for multi-objective optimization of the cloud service providers' workload distribution problem. SHIELD was shown to generate better solutions and generate them faster than other frameworks. In our experiments, SHIELD was able to realize 34.4× speedup and 2.1× improvement in PHV while reducing the carbon footprint by up to 3.7×, water footprint by up to 1.8×, energy costs by up to 1.3×. and a cumulative improvement across all objectives (carbon, water, cost) of up to 4.8×, compared to state-of-the-art frameworks.

## REFERENCES

[1] Amazon, "Global Infrastructure," Amazon, [Accessed 1 May 2023], Available: http://aws.amazon.com/about-aws/global-infrastructure/.

[2] E. Masanet, et al., "Recalibrating global data center energy-use estimates," *Science,* vol. 367, no. 6481, pp. 984-986, 2020.

[3] IEA, "Data Centres and Data Transmission Networks," [Accessed 1 May 2023], Available: https://www.iea.org/reports/data-centres-and-data-transmission-networks.

[4] B. Daigle, "Data Centers Around the World: A Quick Look," *United States International Trade Commission,* 2021.

[5] Google, "Google Data Centers 2021 Annual Water Metrics," 2021.

[6] T. E. Wirth, "The future of energy policy," *Foreign Affairs,* 2003.

[7] F. Libertson, et al., "Data-Center infrastructure and energy gentrification: perspectives from Sweden," *SSPP,* vol. 17, 2021.

[8] Y. Li, H. Wang, et al., "Operating cost reduction for distributed internet data centers," *IEEE/ACM CCGRID,* 2013.

[9] J. Ni, et al., "A review of air conditioning energy performance in data centers," *Renewable And Sustainable Energy Reviews,* 2017.

[10] ASHRAE, "Equipment thermal guidelines for data processing environments," 2021.

[11] N. Hogade, et al., "Minimizing energy costs for geographically distributed heterogeneous data centers," *IEEE TSUSC,* 2018.

[12] T. A. Ndukaife, et al., "Optimization of water consumption in hybrid evaporative cooling air conditioning systems for data center cooling applications," *Heat Transfer Engineering,* vol. 40, no. 7, 2019.

[13] S. Ruth, "Reducing ICT-related carbon emissions: an exemplar for global energy policy," *IETE technical review,* vol. 28, no. 3, 2021.

[14] B. K. Dewangan, et al., "Extensive review of cloud resource management techniques in industry 4.0: Issue and challenges," *Software: Practice and Experience,* vol. 51, 2021.

[15] S. K. Garg, et al., "SLA-based virtual machine management for heterogeneous workloads in a cloud," *J. Netw. Comput. Appl.,* 2014.

[16] H. Yuan, et al., "Temporal task scheduling with constrained service delay for profit maximization in hybrid clouds," *IEEE T-ASE,* 2017.

[17] D. Alsadie, et al., "Dynamic resource allocation for an energy efficient VM architecture for cloud computing.," *ACM ACSW,* 2018.

[18] H. Liu, et al., "A holistic optimization framework for mobile cloud task scheduling," *IEEE TSUSC,* 2017.

[19] J. Bi, et al., "Green energy forecast-based bi-objective scheduling of tasks across distributed cloud," *IEEE TSUSC,* 2021.

[20] H. Tamaki, et al., "Multi-objective optimization by genetic algorithms: A review," *IEEE ICEC,* 1996.

[21] K. Sindhya, et al., "A hybrid framework for evolutionary multi-objective optimization," *IEEE TEVC,* vol. 17, no. 4, 2012.

[22] J. D. Moore, et al., "Making scheduling "cool": temperature-aware workload placement in data centers," *USENIX ATC,* 2005.

[23] R. F. Sullivan, "Alternating cold and hot aisles provides more reliable cooling for server farms," *Uptime Institute,* 2000.

[24] H. M. Daraghmeh, et al., "A review of current status of free cooling in datacenters," *Applied Thermal Engineering,* vol. 114, 2017.

[25] K. M. U.Ahmed, et al., "A review of data centers energy consumption and reliability modeling," *IEEE Access,* vol. 9, 2021.

[26] Q. Zhang, et al., "A survey on data center cooling systems: Technology, power consumption modeling and control strategy optimization," *Journal of Systems Architecture,* vol. 119, 2021.

[27] D, Azevedo, et al., "Water usage effectiveness (WUE): A green grid datacenter sustainability metric," *The Green Grid,* 2011.

[28] M.A. B. Siddik, et al., "The environmental footprint of data centers in the United States," *Environmental Research Letters,* vol. 16, 2021.

[29] P. Torcellini, et al., "Consumptive water use for US power production," *NREL,* 2003.

[30] S., Mohammad, et al, "Carbon dioxide separation from flue gases: a technological review emphasizing reduction in greenhouse gas emissions," *Sci. World J.,* 2014.

[31] U. S. E. I. A. (EIA). [Accessed 1 May 2023], Available: https://www.eia.gov/tools/faqs/faq.php?id=74&t=11.

[32] P. A. Malinowski, et al., "Energy-water nexus: Potential energy savings and implications for sustainable integrated water management in urban areas from rainwater harvesting and gray-water reuse," *J. Water Resour. Plan. Manag,* 2015.

[33] J. Kumar, "BiPhase adaptive learning-based neural network model for cloud datacenter workload forecasting," *Soft Computing,* 2020.

[34] D. G. Feitelson, at al., "Experience with using the Parallel Workloads Archive," *JPDC,* vol. 74, pp. 2967-2982, 2014.

[35] S. Ghanbari, et al., "A priority based job scheduling algorithm in cloud computing," *Procedia Engineering,* vol. 50, 2012.

[36] Q. Zhang and H. Li, "MOEA/D: A multiobjective evolutionary algorithm based on decomposition," *IEEE TEVC,* vol. 11, 2007.

[37] J, Andrzej, "On the performance of multiple-objective genetic local search on the 0/1 knapsack problem-a comparative experiment," *IEEE Trans. Evol. Comput.,* vol. 6, no. 4, pp. 402-412, 2002.

[38] J. A. Boya, et al., "Learning evaluation functions to improve optimization by local search," *J. Mach. Learn. Res.,* 2001.

[39] K. Sindhya, et al., "A hybrid framework for evolutionary multi-objective optimization," *IEEE TEVC,* vol. 17, no. 4, 2013.

[40] M. Zaharia, et al., "Improving mapreduce performance in heterogeneous environments," *Osdi,* vol. 8, no. 4, 2008.

[41] B. K. Joardar, et al., "Learning-Based application-agnostic 3d noc design for heterogeneous manycore systems," *IEEE TC (68),* 2019.

[42] "BigDataBench 5.0 benchmark suite," [Accessed 1 May 2023], Available: http://www.benchcouncil.org/BigDataBench/.